**Path to a Single-Stage, 100-GeV Electron Beam via a Flying-Focus-Driven Laser-Plasma Accelerator**

J. L. Shaw[1,*], M. V. Ambat[1], K. G. Miller[1], R. Boni[1], I. LaBelle[1], W. B. Mori[1], J. J. Pigeon[1], A. Rigatti[1], I. Settle[1], L. Mack[1], J. P. Palastro[1], and D. H. Froula[1]

**Affiliations**

[1]University of Rochester Laboratory for Laser Energetics, Rochester New York 14623
* Corresponding authors: jshaw05@lle.rochester.edu

**Abstract:**

Dephasingless laser wakefield acceleration (DLWFA), a novel laser wakefield acceleration concept based on the recently demonstrated "flying focus" technology, offers a new paradigm in laser-plasma acceleration that could advance the progress toward a TeV linear accelerator using a single-stage system without guiding structures. The recently proposed NSF OPAL laser facility could be the transformative technology that enables this grand challenge in laser-plasma acceleration. We review the viable parameter space for DLWFA based on the scaling of its performance with laser and plasma parameters, and we compare that performance to traditional laser wakefield acceleration. These scalings indicate the necessity for ultrashort, high-energy laser architectures such as NSF OPAL to achieve groundbreaking electron energies using DLWFA. Initial results from MTW-OPAL, the platform for the 6-J DLWFA demonstration experiment, show a tight, round focal spot over a distance of 3.7 mm. New particle-in-cell simulations of that platform indicate that using hydrogen for DLWFA reduces the amount of laser light that is distorted due to refraction at ionization fronts. An experimental path, and the computational and technical design work along that path, from the current status of the field to a single-stage, 100-GeV electron beam via DLWFA on NSF OPAL is outlined. Progress along that path is presented.

**I. Introduction**

High-Energy Physics (HEP) colliders provide a window into the basic building blocks of the universe, but the field relies on colliders with ever-increasing particle energies. Currently, the field of HEP is advocating for particle collisions with center-of-mass energies ~ 10 TeV to access the boundaries of the understandable universe [1]. To date, HEP has used conventional radiofrequency (RF) accelerators. However, as the energy gain from conventional RF accelerator technology begins to plateau, advanced accelerator concepts become imperative to push particle energies to new levels.

In 1979, Tajima and Dawson published the seminal *Laser Electron Accelerator* manuscript [2] that opened the field of laser-plasma acceleration (LPA) and the potential for TeV electron beams. About a decade later, the first electrons from a plasma accelerator were measured [3,4], and in 2004, the Dream Beam publications demonstrated quasi-monoenergetic, MeV electron beams accelerated in millimeter-long plasmas [5-7]. Although many more advancements are needed to achieve the high-quality electron beams required for HEP studies, the recent 5-10 GeV experimental results [8-11] are reaching the necessary conditions for current advanced accelerator concepts, where 100-1000 separate 10-GeV laser wakefield accelerator (LWFA) stages are envisioned to produce the next-generation TeV linear collider [12]. Extending traditional LWFA stages beyond 10 GeV requires lowering the plasma density and extending the laser pulse durations while maintaining relativistic intensities, which is challenging given the available laser energies in current laser architectures. Dephasingless LWFA (DLWFA) driven by an ultrafast flying focus [13, 14] is an original concept that is a disruptive technology with the potential to transform the field of LPA and, more broadly, advanced accelerators.

This manuscript is organized as follows. Section II presents the background of DLWFA and how the concept has been selected as a flagship experiment to guide the design of the proposed NSF OPAL laser user facility [15]. In Section III, we compare the performance of DLWFA to traditional LWFA in terms of laser and plasma parameters; this shows how laser architectures like those of NSF OPAL are necessary for TeV-class DLWFA. An experimental path, including the technical design and computational work, based on four stages of laser development is described in Section IV. New simulations for the 6-J demonstration phase of DLWFA and initial experimental results from that phase are also included in Section IV.

## **II. Background**

In a traditional LWFA, an ultrashort, intense laser pulse propagates through either a neutral gas or plasma. The ponderomotive force of the drive laser expels electrons out and around the front of the laser pulse. On this timescale, the massive ions remain relatively immobile, so an ion column forms behind the drive laser. The expelled electrons are drawn back toward the laser axis by the Coulomb force of the ion column, causing them to overshoot the laser axis and generate a periodic wake structure. The resulting charge separation produces a longitudinal electric field that travels at the group velocity of the drive laser. Electrons that become trapped in the wake can be accelerated by that electric field and reach relativistic energies.

The maximum electron energy from a single-stage, traditional LWFA is constrained by the low plasma densities required to limit dephasing between the laser pulse and accelerated electrons. Electron energy gained in an LWFA is limited by one of three phenomena: diffraction, depletion, or dephasing. Methods such as self-guiding [16] and guiding structures [17, 18] have been developed to control diffraction. Depletion can be circumvented by matching the depletion length to the dephasing length [19] and by utilizing a laser driver with enough energy for its pulse duration. Dephasing occurs when the accelerating electrons rapidly approach the speed of light c and therefore outrun the accelerating portion of the wake, which travels at the group velocity ($v_g < c$) of the drive laser in the plasma. In a traditional LWFA, dephasing places a limit on the length over which electrons can be accelerated, known as the dephasing length. To extend the acceleration length, lower plasma densities have been used to increase the group velocity of the drive laser at the expense of the acceleration gradient.

The ultrafast flying focus is a spatiotemporal focusing system that offers a new paradigm in LPA by using a single-stage system without guiding structures [14]. The ultrafast flying focus provides the ability to propagate a high-intensity laser pulse over meters at a custom velocity close to c while maintaining a small focal spot and a near-transform-limited pulse duration. By controlling the velocity of a focal spot, a wakefield can be driven at the speed of light in a plasma, thus eliminating dephasing. Decoupling the velocity of the wakefield from the plasma density and removing the need to guide the laser over long distances removes two significant constraints in traditional LWFA and enables a DLWFA that could provide electron beam energies limited only by the available laser energy in a single stage. Relieving these constraints also allows the LPA to be optimized in original ways, opening new opportunities for improving electron beam quality (e.g., energy spread, peak current, transverse emittance, etc.). Initial scalings [20] of this "dephasingless" LWFA suggest that a 20-fs, 500-J laser (NSF OPAL) would be capable of accelerating electrons to 125-GeV energies in a single, one-meter stage. Presently in its design phase, NSF OPAL is a next-generation user facility that will provide the highest-intensity laser driver in the world (two 25 PW beams—500 J in 20 fs) to enable experiments at the power and intensity frontiers [15]. With its high energy in an ultrashort pulse duration, the proposed NSF OPAL facility could open the path to a single-stage, TeV-class electron beam via flying-focus-driven LPA. This remarkable performance compares with the proposed International Linear Collider that would be between 30- and 50-km long (if constructed) at an estimated cost of $6.7 billion (in 2007 US dollars, excluding R&D, prototyping, land acquisition, underground easement costs, detectors, contingencies, and inflation), per its Reference Design Report [21].

This concept of a flying-focus-driven LPA has been selected as a flagship experiment to guide the design of NSF OPAL. The ultimate goal of this flagship experiment is to demonstrate 100-GeV electron energies from a single plasma stage using an ultrafast flying focus driven by NSF OPAL. The results of this flagship could provide a step-wise change in the direction of the field of advanced accelerators and eventually be a gamechanger for HEP. A successful 100-GeV flagship could shift LPA laser drivers to modern, high-power laser architectures, given that optimizing for energy gain requires short laser pulses and high laser energies as described in Section III. Additionally, demonstrating 100-GeV energies in a single stage could shift the emphasis of the LPA field away from staging hundreds of 10-GeV stages to reach TeV energies.

The ultrafast flying focus uses spherical aberration to create an extended focal region and a radial echelon to introduce radial group delay that controls the time at which rings of power reach their respective foci, as shown in Figure 1. The shape of the echelon provides control over the trajectory. The extended focal range is generated by the axiparabola [22], which is a parabola with intentional spherical aberration. This optic combination allows the laser pulse duration to remain short and the focal velocity to be decoupled from the plasma conditions, allowing an ultrashort, high-intensity laser pulse to drive a wakefield at a custom velocity. Note that an ultrashort, ultrafast flying focus can be produced with an axiparabola alone, but it has an uncontrolled and accelerating focal velocity; DLWFA requires control of the focal velocity.

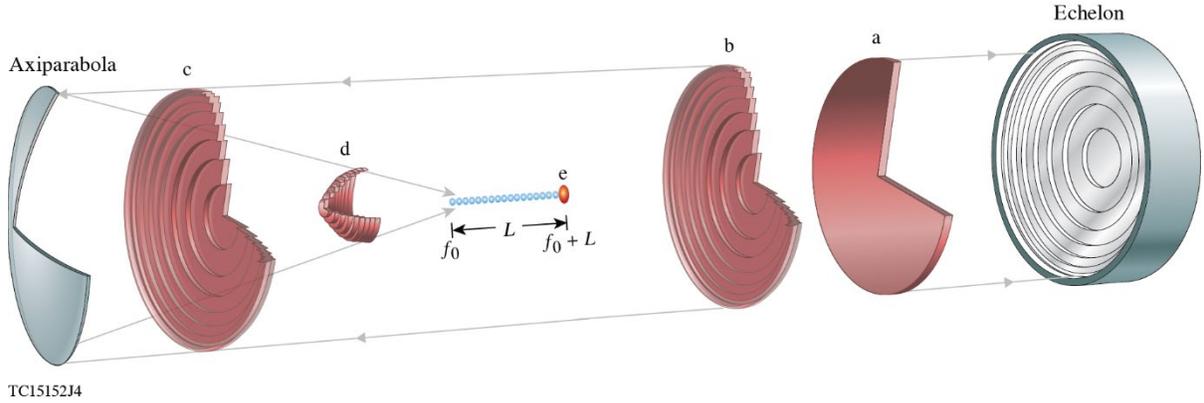

**Figure 1: Visual representation of the formation of an ultrafast flying focus for DLWFA by reflection from (a,b) a radially stepped echelon followed by (c) an axiparabola. (d) Different radii in the near-field are focused to different locations in the far-field. The location of the initial focus is $f_0$, and L is the distance of the extended focus over which the light is spread. (e) The ultrafast flying focus pulse can drive a wake at the speed of light. Reprinted figure with permission from J. P. Palastro, J. L. Shaw, P. Franke, D. Ramsey, T. T. Simpson, and D. H. Froula, Phys. Rev. Lett. 124, 134802 (2020). Copyright 2020 by the American Physical Society.**

If dephasing in LPAs could be overcome, electrons could be accelerated to much higher energies over the same length. In a traditional LWFA, the energy gain $\Delta E_{trad}$ is proportional to the acceleration length $L_{trad}$: $\Delta E_{trad} \propto k_p L_{trad} \sqrt{a_0}$ where $k_p = \omega_p/c$, $\omega_p = \sqrt{\frac{e^2 n_e}{m\epsilon_0}}$ is the plasma frequency, and $a_0$ is the normalized vector potential $a_0 \cong 8.5 \times 10^{-10} \sqrt{I_0[W/cm^2]} \lambda[\mu m]$. Here, e is the electron charge, $n_e$ is the plasma electron density, m is the electron mass, $\epsilon_0$ is the permittivity of free space, $I_0$ is the drive laser intensity, and $\lambda$ is the wavelength of the drive laser. The acceleration length scales as $L_{trad} \propto \frac{\omega_0^2}{\omega_p^3} \sqrt{a_0}$, where $\omega_0 = c/2\pi\lambda$. This means that lower plasma densities provide longer acceleration length and larger electron energies given that $\Delta E_{trad} \propto n_e^{-1}$. Electron energy gain in DLWFA, $\Delta E_{DLWFA}$, is also proportional to the acceleration length: $\Delta E_{DLWFA} \propto k_p L_{DLWFA} \sqrt{a_0}$. Here, however, the acceleration length $L_{DLWFA}$ is not limited by dephasing but is only constrained by the laser energy and technology. This behavior allows for operation at higher plasma densities, which provide larger electron energies given that $\Delta E_{DLWFA} \propto n_e^{1/2}$. Figure 2 shows the impact of this scaling. The accelerating gradient of traditional LWFA is orders of magnitude higher than for conventional RF accelerators, leading to much higher electron energies in the same acceleration length; however, the ability of DLWFA to operate at even higher plasma densities further increases the possible electron energy achievable in the same acceleration length.

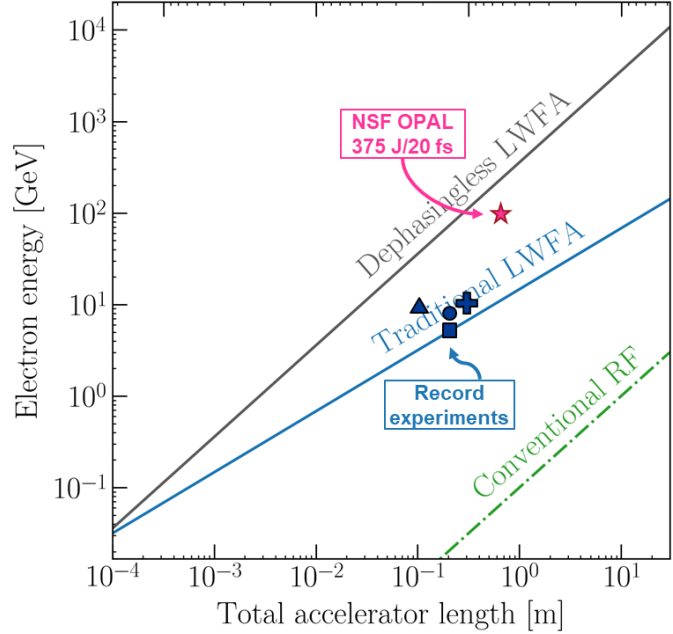

**Figure 2:** Electron energy gain versus total accelerator length for three technologies: conventional RF accelerators (green dot-dash), traditional LWFA (blue), and DLWFA (grey). The pink star is the prediction for the electron energy of a DLWFA produced by NSF OPAL (apodized to 375 J in 20 fs). The blue symbols mark the best experimental results to date from a traditional LWFA (square from Ref. [8], circle from Ref. [9], triangle from Ref. [10], cross from Ref. [11]). Note that the calculation for the DLWFA curve assumes a drive laser pulse duration of 15 fs FWHM.

Overcoming dephasing has been a topic of interest in LPA for many decades. Early work [23, 24] investigated the use of tapered channels to extend interaction distances. Simulations investigating quasi-phase matching (QPM) in corrugated plasma channels showed that QPM could enable electron acceleration past the dephasing length [25]. 2D simulations demonstrated a concept where the plasma density profile was modulated to create resonant LWFA sections followed by nonresonant drift sections that allow the electrons to rephrase [26]. Debus *et al*. demonstrated in simulations a concept utilizing the overlap of two obliquely incident laser pulses with tilted pulse fronts to drive a wake with the same trajectory as the accelerated electrons [27].

The first concepts for spatio-temporal control to overcome dephasing arrived in the late 2010s. Concurrent with the development of the flying focus at LLE [14, 20], C. Caizergues *et al*. [28] showed in simulations that the utilization of pulse front delay coupled with an axiparabola can lead to LPA in a dephasing-free regime where the electron energy gain is higher by more than an order of magnitude. In parallel with LLE, the experimental team from the Weizmann Institute of Science and Laboratoire d'Optique Appliquée have been pioneering a spatio-temporal shaping concept based on an axiparabola and a refractive doublet that controls the pulse front curvature. Initial characterizations of the focal capabilities of an axiparabola were demonstrated [22], and they then showed the first measurements of superluminal and subluminal focal trajectories using their method in 2024 [29]. Recently, they measured the first plasma wakefield structures based on this form of spatio-temporal shaping [30].

# III. Design of DLWFA with an ultrafast flying focus

The efficiency of a DLWFA is primarily driven by the laser pulse duration relative to the plasma density. For laser pulse durations less than 21.1 fs, DLWFA uses less laser energy and is more efficient than traditional LWFA (see in Figure 3) because ultrashort laser drivers enable operation at higher plasma density, which yields larger accelerating gradients. These performance scalings mean that DLWFA is ideally suited to the current trend in new laser architectures, such as NSF OPAL, with ultrashort pulse durations and large energy. Note that while the efficiency is better for DLWFA using short pulse durations, DLWFA operating over a wide range of pulse durations outperforms LWFA in terms of maximum electron energy for long accelerators, as discussed in the next paragraph.

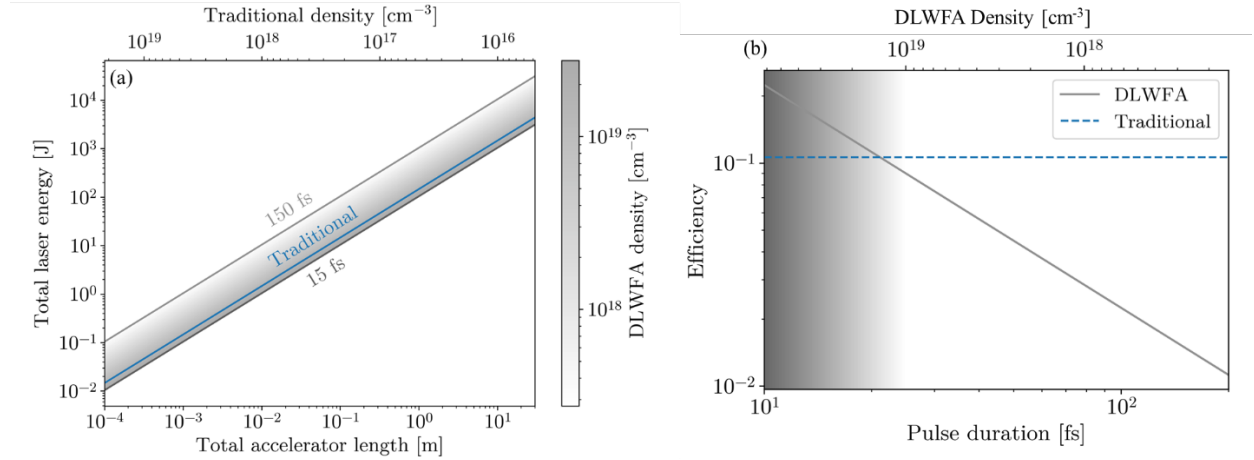

**Figure 3: (a) Required laser energy versus total accelerator length for traditional LWFA (blue curve) and DLWFA with drive laser pulses of varied duration (grey curves). The color bar indicates the operating plasma density of the DLWFA. The operating plasma density of the traditional LWFA is given by the top axis. (b) Efficiency versus pulse duration of the laser driver for traditional LWFA (blue dashed curve) and DLWFA (solid grey curve). The greyscale implies the increasing difficulty of producing pulses with durations less than ~20 fs.**

Like Figure 2, Figure 4(a) shows the accelerated electron energy versus acceleration length. Added in Figure 4(a), however, is a range of electron energies at a given acceleration length for DLWFA (region bounded by the grey curves) that depends on the pulse duration of the drive laser. The shorter the drive laser, the higher the plasma density, and therefore the higher the gradient, at which the DLWFA can operate. This yields greater energy gains in an equivalent length. On Figure 4(a), the curve for traditional LWFA employing 100 stages [31] is also included, which still underperforms DLWFA for all pulse durations shown in reaching electron energies above 1 TeV. For both a traditional LWFA and a DLWFA, the amount of charge that can be accelerated depends on the size of the wake, and thus the operating plasma density of the LPA. For a DLWFA, the plasma density is set by the pulse duration of the drive laser, and so there is no variation of charge with the total accelerator length, as can be seen in Figure 4(b). Conversely, because the total acceleration length of a traditional LWFA depends on the plasma density, the charge also varies with the total acceleration length. Note that the scalings in Figure 3 and Figure 4 assume $a_0 = 2$ (matched for LWFA [19]), and the methodology to produce the scalings is included in the Appendix.

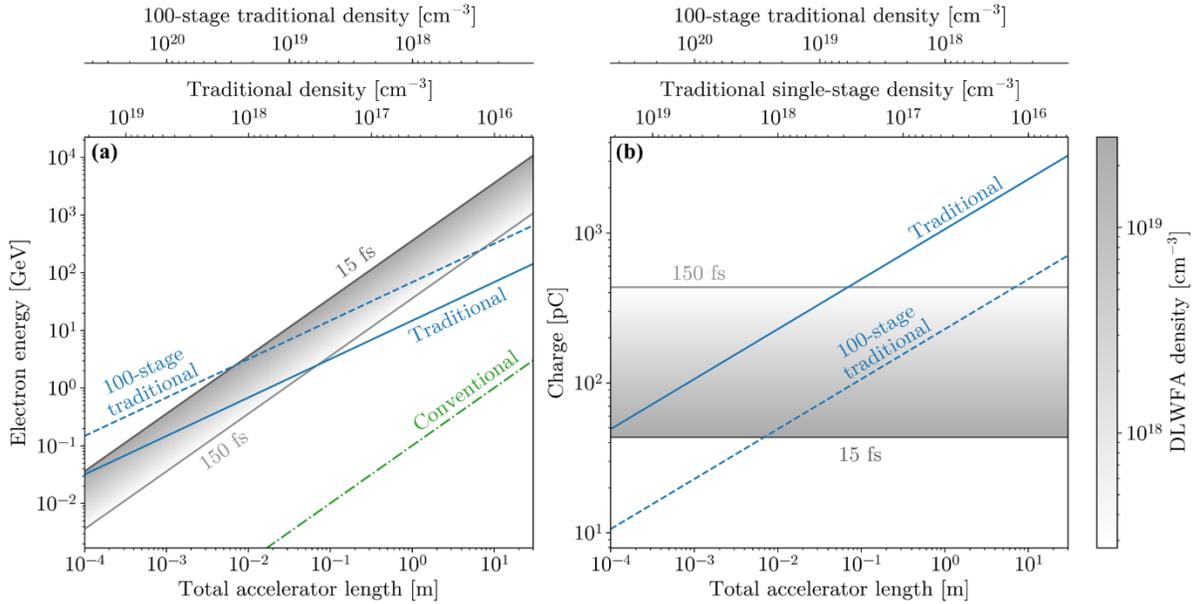

**Figure 4:** (a) Electron energy gain versus total accelerator length for three technologies: conventional RF accelerators (green dot-dash), traditional LWFA (blue), and DLWFA (grey). (b) Accelerated charge versus total accelerator length. The curves for DLWFA (grey) are labelled by the duration of the drive pulse. For both plots, the color bar indicates the plasma density for the DLWFA that corresponds to the given pulse duration. The blue curves show the charge for a traditional LWFA (solid curve) and a 100-stage traditional LWFA (dashed curve). The plasma density for the traditional LWFA and 100-stage traditional LWFA is given by the scales at the top of both plots.

Because DLWFA is predicted to provide electron beam energies limited only by the available laser energy, the research is ideally suited for the highest-energy laser that exhibits an ultrashort pulse duration. The NSF OPAL design calls for 500 J in 20 fs [15], which is more laser energy than is available anywhere with an ultrashort pulse duration. Therefore, it is uniquely suited to demonstrate a 100-GeV DLWFA. Note that the current NSF OPAL design is 500 J in a square beam. Current understanding is that a round beam is required for DLWFA with an axiparabola-echelon pair. Apodizing NSF OPAL from a square beam to a round beam reduces the available energy to 375 J and thus reduces the predicted electron energy from the DLWFA to 100 GeV. This energy will be used for the remainder of manuscript. Additionally, NSF OPAL will feature a "Beta" beam that can be run at 40 J (30 J if apodized round), but at a higher repetition rate [15]. The Vulcan 20-20 upgrade design [32] calls for 400 J in 20 fs, so 100-GeV DLWFA could also be completed there if proper experimental space was allocated. Other existing laser facilities with the required ultrashort pulse duration (<30 fs), such as ZEUS [33], CoReLS [34], and L3 at ELI Beamlines [35], do not offer such energies. Therefore, these facilities are useful stepping stones to the flagship experiment, but ultimately are not expected to be able to produce 100-GeV electron beams.

## IV: Path to Flagship

The ultimate goal of the flagship experiment is to demonstrate 100-GeV electron energies from DLWFA in a single stage using NSF OPAL. We have laid out a systematic plan of computational, technical design, and experimental milestones aimed at scaling DLWFA with increasing laser energies and capabilities. With the completion of the demonstration of the ultrafast flying focus with the broadband, super-luminescent diode (Phase 1) [36], the plan targets three additional phases designed to ramp up in parallel with the NSF OPAL project: Phase 2—DLWFA with MTW-OPAL (6-J demonstration); Phase 3—DLWFA with the NSF OPAL Beta Beam (30 J/20 fs demonstration); and Phase 4—Flagship with the full NSF OPAL (375 J/20 fs). In each subsection below, the computational, technical design, and experimental work required to complete each phase is outlined.

## IVa: Phase 1—CW Demonstration of an ultrafast flying focus

**Experiment**

At LLE, an all-optical demonstration of an ultrafast flying focus using a broadband, superluminescent diode demonstrated the key features that could enable the system to be used for DLWFA [36]. Figure 5(a) and Figure 5(b) show that the focal velocity can be controlled by utilizing appropriate axiparabola-echelon pairs. For the case using the convex echelon [Figure 5(a)], the measured velocity is subluminal (~0.9987c); the concave echelon [Figure 5(b)] produces a superluminal velocity (~1.002c). For both cases, the focal region extents for more than 50 Rayleigh ranges. This work demonstrated that a controllable, ultrabroadband flying focus with a diffraction-limited [Figure 5(c)] and nearly transform-limited [Figure 5(d)] moving focus can be created over a centimeter scale. This work additionally verified that we could fabricate the necessary optics to produce an ultrafast flying focus.

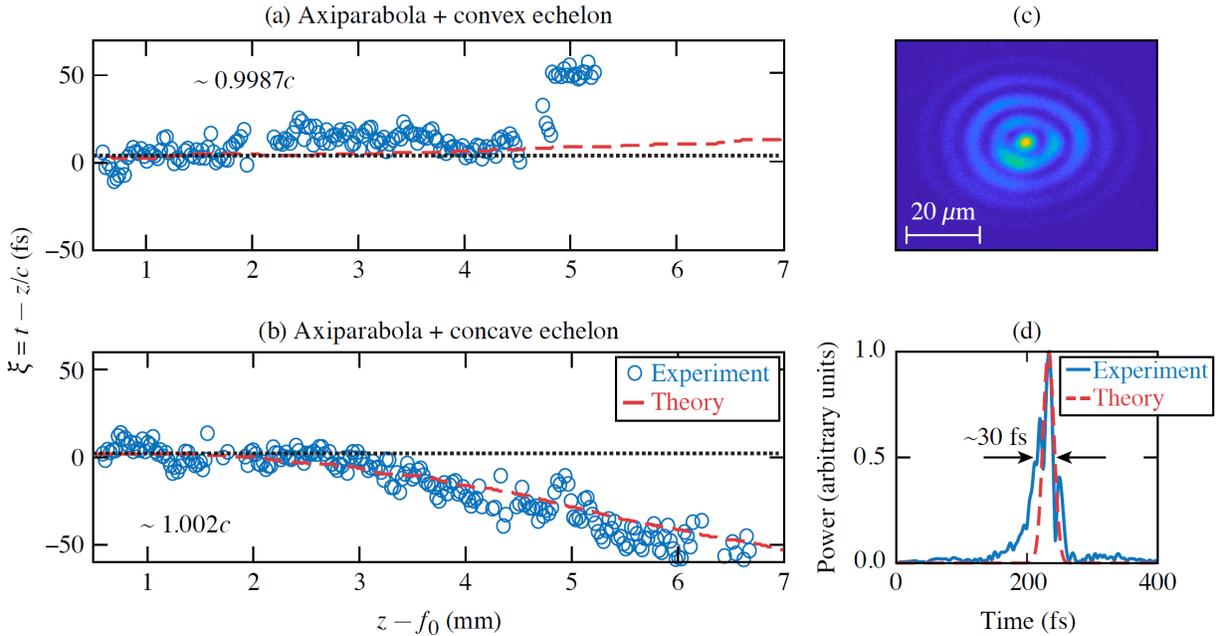

E30886J2

**Figure 5:** Focal velocity measurements for an axiparabola and an (a) convex or (b) concave echelon. The delay data is plotted as a function of the speed-of-light variable $\xi$ relative to the nominal focus $f_0$. (c) Spatial beam profile. (d) Effective pulse duration measurement. Reprinted with permission from [36] © Optica Publishing Group.

**Technical Design**

Significant technical development was required to enable the demonstration of the ultrafast flying focus. LLE developed an in-house capability using electron-beam evaporation to fabricate the echelons [Figure 6(a)] that produce an ultrafast flying focus based on the designed step profile for the application [Figure 6(b)]. White light interferometry measurements [Figure 6(c-e)] ensure that the manufactured echelons meet the specifications for experiments. LLE has also fabricated axiparabolas with this method, and they are now in use on THz-producing flying focus experiments [37].

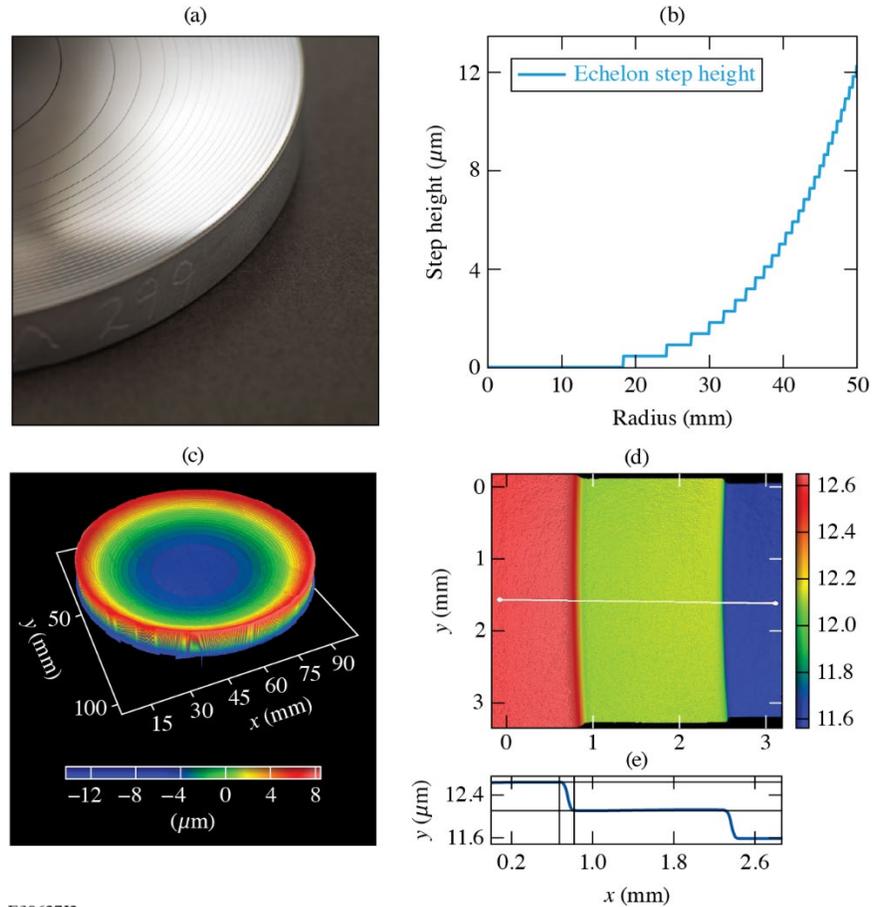

E30627J2

**Figure 6: (a) Photograph of an LLE-fabricated echelon. (b) Design of echelon step height for fabrication. (c) White light interferometry measurements of the fabricated step profile. (d) Zoomed in version of (c) in the vicinity of a single, half-wavelength step. (e) Profile of (d) as indicated by the white line. (b-e) are adapted with permission from [36] © Optica Publishing Group.**

There are ongoing efforts to develop new, novel concepts [38, 39] to produce flying focus pulses that would also allow for DLWFA. New concepts include a dynamic flying focus that is based on replacing the echelon with a deformable mirror (DM) and spatial light modulator (SLM) [38] as shown in Figure 7. In this concept, the DM adds programmable radial group delay to control the trajectory. The SLM locally flattens the phase fronts after the DM via phase delay followed by diffraction. This concept would allow programmable velocity for active feedback and machine learning, and it has recently been demonstrated [40]. As this concept and other concepts mature, they may further simplify the production of flying-focus optics.

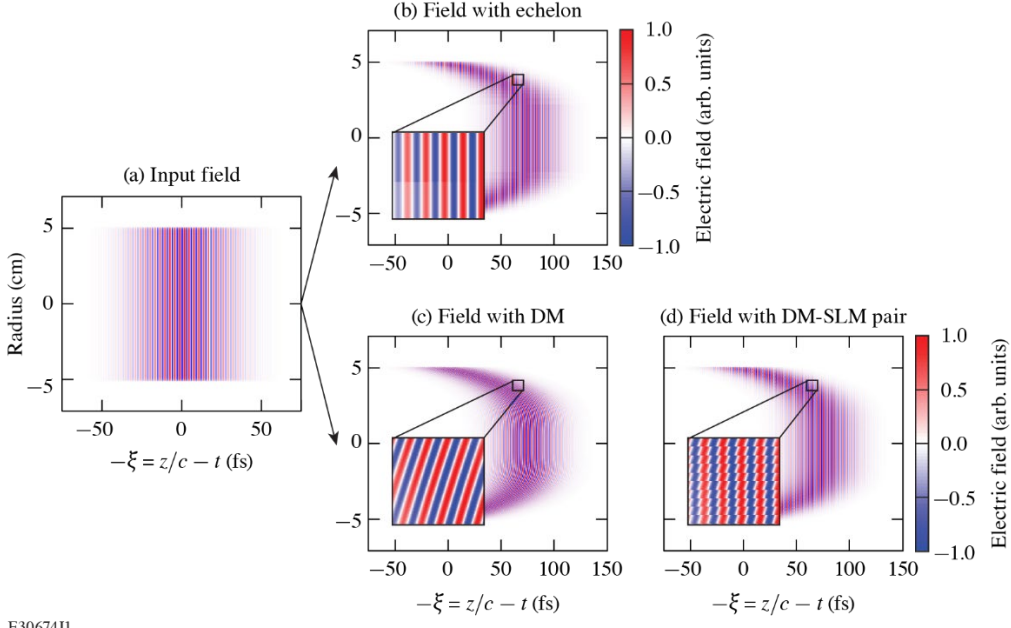

**Figure 7:** Concept of a programmable flying focus using a DM-SLM pair shown through the modification to the electric field in the near-field. (a) Input laser field with an ultrashort pulse duration, flat phase fronts, and a flat pulse front. (b) Field structure after reflection from an echelon, as is done in the ultrafast flying focus. The echelon imparts designed radial group delay while keeping the phase fronts flat. (c) Field after DM. Appropriate radial delay is applied, but phase fronts are curved. (d) Field after DM-SLM pair. SLM locally flattens the phase fronts via phase delay followed by diffraction. Note that phase fronts are globally flat but have a residual tilt within each pixel. Reprinted with permission from [38] © Optica Publishing Group.

### IVb: Phase 2—DLWFA with MTW-OPAL (6-J Demonstration)

The path to flagship is currently in Phase 2, where we are seeking to demonstrate DLWFA with MTW-OPAL, which is the prototype for NSF OPAL and therefore uses the same technologies and addresses many of the relevant challenges.

**Computation**

Before experiments began, the ability to model flying focus pulses produced by an axiparabola-echelon pair was implemented into the particle-in-cell (PIC) code OSIRIS [41] to simulate DLWFA. Since then, a generalized mathematical method for generating spatio-temporally shaped pulses has been developed and was demonstrated in OSIRIS [42]. OSIRIS simulations using MTW-OPAL-relevant parameters demonstrated that a DLWFA driven with an ultrafast flying focus can accelerate 25 pC of charge over 20 dephasing lengths to a maximum energy gain of 2.1 GeV [20], as shown in Figure 8. The average energy spread was 1.8%, and the laser-to-electron-beam efficiency was 0.7%. The drive laser pulse had a central wavelength of 1054 nm, 6.2 J of energy, and a pulse duration of 15 fs. The modeled axiparabola was f/7 with an initial focus $f_0$ of 0.7 m and a 5-cm optic radius. It produced an extended focal region L of 2 cm. That combination produced an $a_0$ of 3 in vacuum and 2 in the plasma. The plasma had a plateau density of $7 \times 10^{18}$ cm$^{-3}$ that was 2 mm in length and was comprised of 90% He/10% Ar (pre-ionized to 8). The argon enabled ionization injection [43, 44].

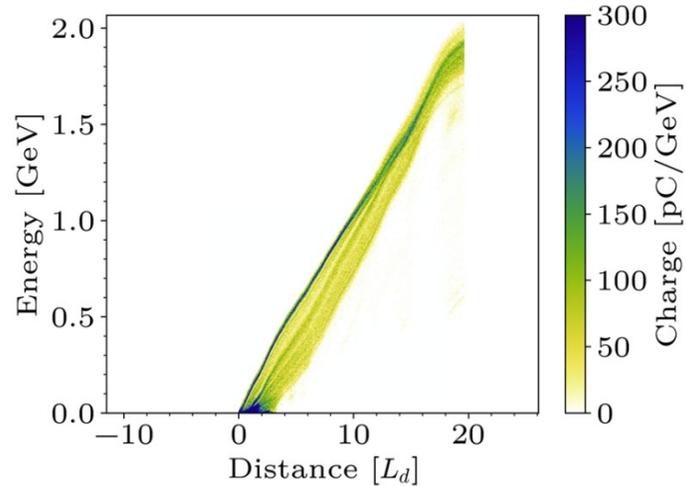

**Figure 8:** Energy gain of the ionization-injected electrons in the first bubble. Material adapted from K. G. Miller *et al.* "Dephasingless laser wakefield acceleration in the bubble regime." Sci. Rep., published 2023 [20].

Whereas the original DLWFA simulations [20] for an axiparabola-echelon pair shown in Figure 8 used a pre-ionized plasma, the need to pre-ionize in experiment is a concern. To address that concern, two scaled-down OSIRIS simulations (7 mm in length) were completed, one in neutral hydrogen and one in neutral helium. The resulting plasma density after full ionization is $7 \times 10^{18}$ cm$^{-3}$. The simulations used the current MTW-OPAL laser parameters: central wavelength of 920 μm, a vacuum pulse duration of 23.4 fs FWHM, and 4.7 J of energy. The axiparabola-echelon pair was modeled as one optic with an $f_0$ of 63 cm, a nominal focal length L of 7.8 mm, and a radius of 4.5 cm with the inner 2.25 cm (50%) apodized. 3.4 J of laser energy remain after the apodization. This particular axiparabola-echelon pair was designed to produce the longest focal range possible with $a_0 > 1$ to investigate pulse propagation effects, rather than to drive DLWFA. The focal velocity was computed to be slightly superluminal in the plasma so that the back of the bubble remains at a constant location as in Ref. [20].

Figure 9 shows that when neutral helium is used, the two ionization fronts cause a large portion of the drive laser energy to refract towards the axis. This refraction creates multiple on-axis intensity peaks, significantly disrupting the accelerating structure. Effects from the first and second ionization fronts [marked by arrows with "1" and "2", respectively, in Figure 9(a)] can be seen in the transverse profile of the laser pulse in Figure 9. The on-axis effect, however, is small early in the focal region, as can be seen in Figure 9(b), where the red curve shows that the majority of the laser contributes to driving the wake. As the DLWFA continues, the ionization fronts refract a portion of the drive laser into focus [at z-$f_0$ ≳ 4.6 mm as marked by the arrows in Figure 9(c,d)] in front of the main pulse, disrupting the structure of the DLWFA. The multiple peaks modify the wake structure enough to result in a loss of accelerated charge before the end of the focal region, reducing the achievable electron energy gain.

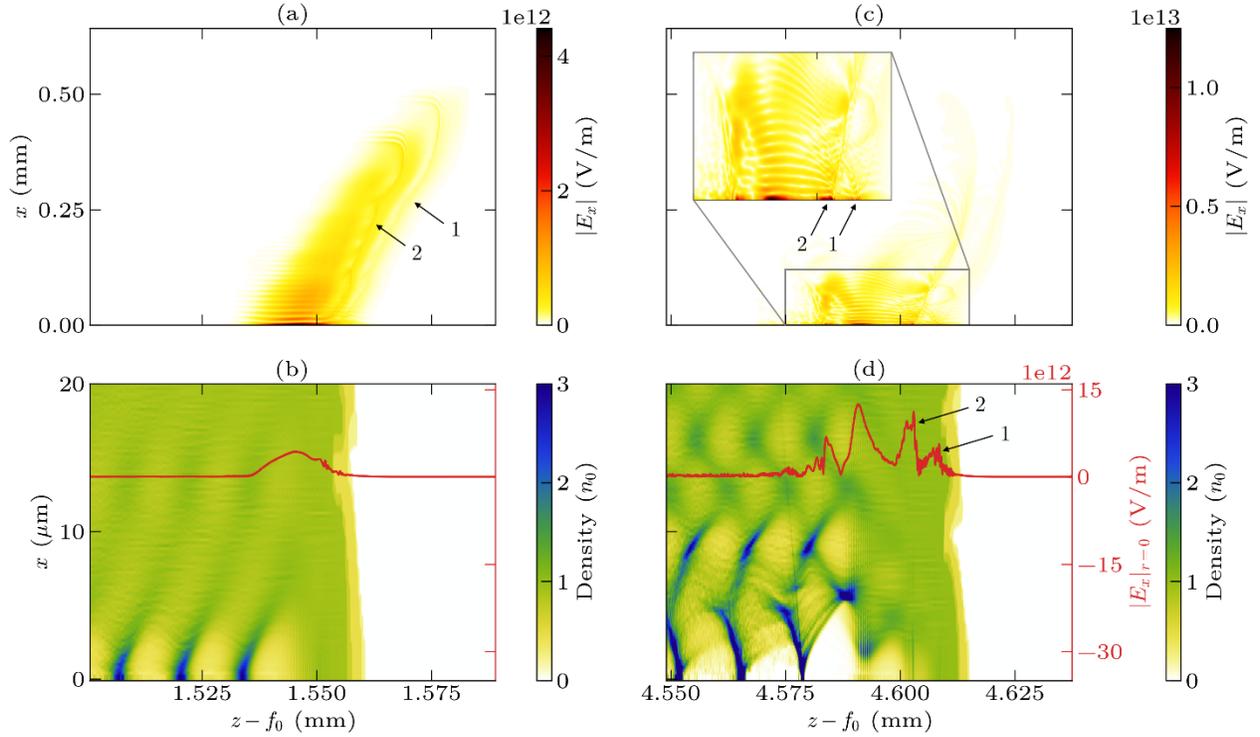

**Figure 9: OSIRIS PIC simulation of DLWFA using neutral helium. The plasma density ramps up over the first 1.7 mm of the focal region. Transverse electric field envelope of the laser at (a) 1.55 mm and (c) 4.60 mm into the focal region. Plasma density at (b) 1.55 mm and (d) 4.60 mm into the focal region. The red curves in (b) and (d) show the on-axis lineout of (a) and (c), respectively. The arrows labelled "1" and "2" mark the results of refraction from the first and second ionization fronts, respectively, in (a, b, d).**

Conversely, DLWFA can be driven for extended distances when starting with neutral hydrogen [Figure 10]. With only one ionization level at a lower energy, there is minimal splitting of the drive laser, as can be seen by comparing Figure 10(a) and (b) in hydrogen to Figure 9(a) and (b) in helium. As a result, later in the plasma, only a small portion of the drive laser focuses in advance of the main bubble [arrow labelled "1" in Figure 10(c,d)] and simply contributes to the ionization. Although these simulations are promising, the possible impact of refraction due to ionization is an open question in DLWFA research.

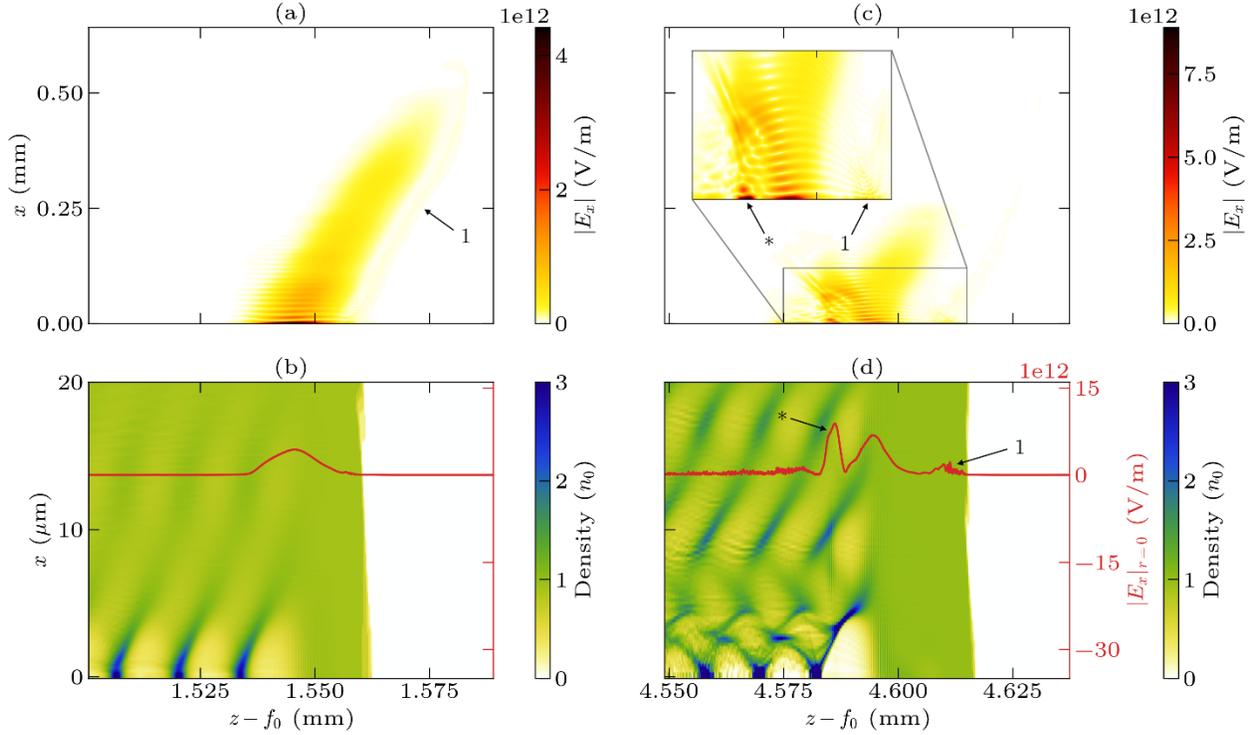

Figure 10: OSIRIS PIC simulation of DLWFA using neutral hydrogen. The plasma density ramps up over the first 1.7 mm of the focal region. Transverse electric field envelope of the laser at (a) 1.55 mm and (b) 4.60 mm into the focal region. Plasma density at (c) 1.55 mm and (d) 4.60 mm into the focal region. The red curves in (c) and (d) show the on-axis lineout of (a) and (b), respectively. The minimal effects of refraction from the hydrogen ionization front are marked with the arrow labeled "1" in (a,b,d). The second main, on-axis peak of light indicated with the arrow labeled * in (c,d) is light trapped in the bubble.

### Experiment

Current experiments are working to demonstrate a DLWFA for the first time using MTW-OPAL [45] operating at 6 J with a pulse duration of 22 fs FWHM. The optics that were characterized in Ref. [36] and used in the OSIRIS simulations in Ref. [20] are currently installed for these experiments. At these laser energies, a traditional gas cell can be used as a plasma source once modified to permit the f/7 cone of the ultrafast flying focus. As the predicted electron energies are moderate (500-800 MeV), a traditional permanent dipole spectrometer is being utilized [46]. This platform was activated over summer 2024, and the extended focal volume produced by an ultrafast flying focus on MTW-OPAL was demonstrated.

Figures 11 (a) and (b) show that the axiparabola can produce a tight, round focal spot over an extended focal range, in agreement with predictions. In these figures, the axiparabola is nominally $f_0 \sim 630$ mm upstream of $z = 0$ mm, and the ultrafast flying focus pulse propagates from left to right, with light rays from smaller near-field radii corresponding to far-field locations closer to the axiparabola (i.e., smaller values of $z$). The axiparabola used in these experiments was designed to produce an extended focal range $L_{des} = 7.8$ mm at a nominal focal length of $f_0 = 630$ mm for an input laser pulse near-field diameter $D = 90$ mm. To model the far-field behavior of the ultrafast flying focus pulse, the near-field spatial fluence profile of MTW-OPAL was measured using a calibrated CCD camera [Figure 11(c)] and served as input to a flying focus model [14, 36] that Fresnel propagates the near-field profile after reflecting off of an axiparabola. Note that Figure 11(c) shows the shadow of an alignment crosshair, which was removed prior to data collection. The same procedure was applied to an ideal near-field spatial fluence profile of MTW-OPAL [Figure 11(d)]. Measurements [blue points in Figure 11(a)] of the normalized peak laser fluence in the far-field accurately capture the rise, two prominent local maxima, and drop-off predicted by modeling of the measured near-field [black curve in Figure 11(a)]. The ripples in the fluence predicted by the model, due to the Bessel-like nature of these ultrafast flying focus pulses, is well represented by the measurements. We measure a focal range $L_{meas}$ ~ 3.7 mm, defined as the length over which the normalized peak fluence is above 0.5, representing the focal distance usable for future high-power applications.

An ideal near-field profile is expected to produce an extended focal range much closer to the design focal range of the axiparabola [grey curve in Figure 11(a)]. The reason for this discrepancy between the focal range produced using the real MTW-OPAL near-field profile versus the ideal MTW-OPAL near-field profile is largely due to the fact that the real MTW-OPAL profile has a shape that more closely resembles a polygon with a maximum side length of ~ 80 mm. This extent is smaller than the axiparabola design laser input diameter of 90 mm; this means that there were fewer light rays at large radii to contribute energy at farther longitudinal locations. The result of this size difference is a shorter focal range.

Measurements of the far-field spot size along its major and minor axes [Figure 11(b)] reveal a high degree of symmetry (to within ~ 1 μm), demonstrating that the axiparabola spot size is round over the ~ 3.7-mm-long focal range. Towards the end of the focal range, the focal spot deteriorates due to exaggerated spherical aberration, owing to these rays corresponding to radii further from the optical axis. Typical focal spot images, shown in Figures 11(e-g), demonstrate that a high-quality focal spot is maintained over the extended focal range.

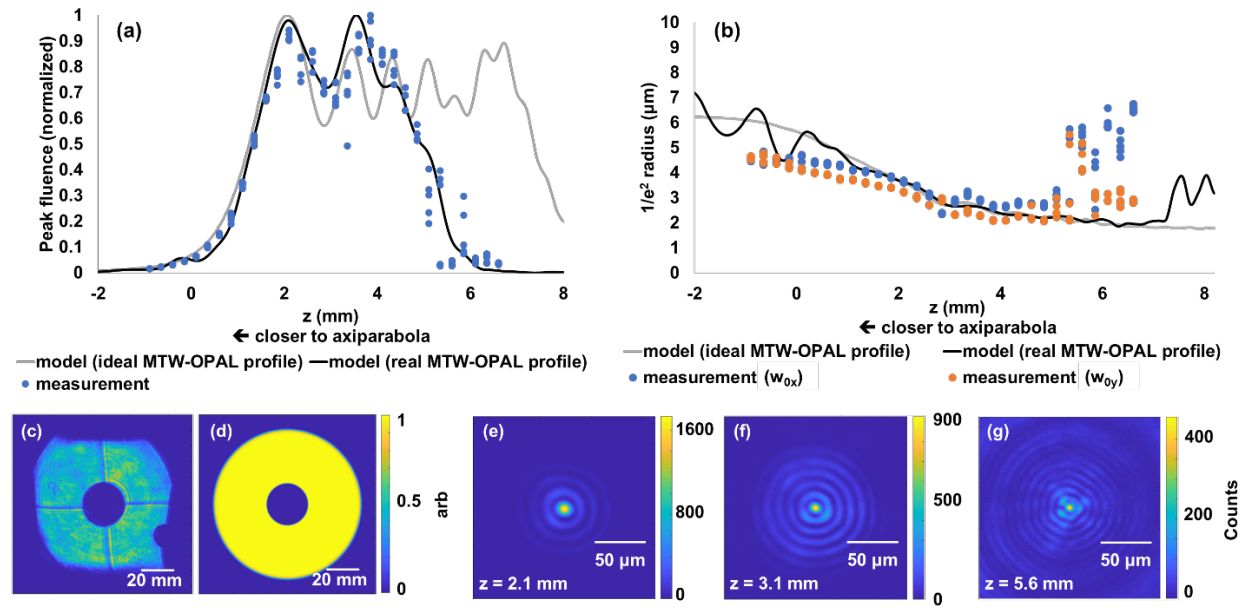

**Figure 11: The axiparabola extended focal range produced by MTW-OPAL. (a) Normalized peak laser fluence and (b) spot size over the length of the extended focal region. (c) Real and (d) ideal MTW-OPAL spatial fluence near-field profiles at the axiparabola surface. (e) – (g) Axiparabola spot size spatial profile measurements at different locations in the far-field extended focal region.**

Beginning in April 2025, a series of campaigns will activate and characterize the plasma source and look for the first demonstration of DLWFA. Subsequent campaigns will optimize the DLWFA and study the fundamentals of particle trapping. The trajectory of the ultrafast flying focus will be optimized to trap a narrow-energy-spread electron beam and accelerate it over distances > 1 cm. This work will study the physical tradeoffs between energy again, total accelerated charge, and beam quality. The findings will be compared against theory and used to inform the next design and experimental steps. The plasma parameters, such as the gas mixture, gas backing pressure, source length, and source position, will also be tuned as needed.

### IVc: Phase 3—Flying Focus Light on NSF OPAL (30-J Demonstration)

To bridge the gap between the 6-J demonstration and the 375-J flagship, the first flying-focus experiments on NSF OPAL will be executed using the round Beta beam (30 J), which is predicted to produce electron beams with energies ~ 7 GeV based on the scaling in Ref. [20].

**Computation**

In preparation for these experiments, 30-J simulations will be conducted. To enable such simulations, flying focus pulses need to be implemented in the quasi-static code QPAD [47], where the focal trajectory and performance can be studied for 10-cm-scale DLWFA. The QPAD results will be verified against OSIRIS for sub-scale runs. In parallel, the capability to model flying-focus pulses in the Lorentz-boosted frame, which shrinks the overall computation load by a factor of ~40, will be added to OSIRIS. Any new physics uncovered by these simulations will be used to steer the experimental campaign plan.

**Technical Design**

A new axiparabola-echelon pair will be designed and fabricated for the 30-J demonstration. In order to properly design these optics, careful characterization of the laser near-field of the NSF OPAL Beta beam is required. The plasma source, which will be scaled up ~ 5x in length from the 6-J design to match the laser energy, will incorporate all lessons learned from the 6-J demonstration and will be fully characterized before experiment. At 7-GeV energies, a standard permanent magnet electron spectrometer such as the ones used in Ref. [8-11] can be used.

**Experiment**

The experimental campaign will target the same goals as the 6-J demonstration and the eventual 375-J flagship experiment: measure the extended focal volume, energy ramp up, acceleration optimization, and trapping control. The exact goals will be informed based on lessons learned from the 6-J demonstration and simulation work. This campaign will also serve to commission the round Beta beam capability on NSF OPAL and the apodizer requirements for the ultrafast flying focus optics. If the LLE DLWFA team is ready for 30-J experiments before the Beta beam is available at NSF OPAL, the study could also be completed on other user facilities with tens of joules and ultrashort laser pulses (CoReLS at 80 J/<20 fs [34] or L3 at Eli Beamlines at 30 J/ <30 fs [35] or ZEUS at 75 J/25 fs [33]).

### IVd: Phase 4—Flying-Focus-Driven LPA for Single-Stage, 100-GeV Electron Beams (Flagship with NSF OPAL Alpha Beam)

The ultimate goal of the flagship experiment is to demonstrate 100-GeV electron energies from a single plasma stage using a flying focus driven by the round NSF OPAL Alpha beam.

**Computation**

Extrapolating the simulations from Ref. [20] to a 375-J, 20-fs-long laser pulse suggests electron energies of 100 GeV in a single stage less than one meter in length. Prior to the 100-GeV flagship experiment, the goal is to complete a full-scale simulation with a 375-J pulse. These simulations are prohibitively expensive with current PIC capabilities due to their nearly 1-m plasma length and the large transverse width required to properly model the flying focus drive laser. To complete these simulations, the capability to model flying- focus pulses in QPAD that was implemented in Phase 3 will be utilized to run full-scale simulations. By combining the Lorentz-boosted frame with injectors from a radial boundary, it may also be possible to simulate the full 1-m stage in OSIRIS.

**Technical Design**

There are three Grand Challenge technologies required for the flagship experiment: meter-scale optics, a meter-scale plasma source, and a method to measure 100-GeV electrons. The fabrication techniques used to make ultrafast flying focus optics need to be extended to the meter-scale apertures required for echelons and axiparabolas for NSF OPAL. A plasma source scaled to 375 J needs to be demonstrated. The primary challenge for the plasma source for the 100-GeV DLWFA flagship is that the plasma source needs to have both a large aperture and a length of ~1 meter. Ultimately, if the full-scale plasma source is not achieved, the sub-scale plasma source would limit the ultimate energy gain of the electrons.

At its core, the 100-GeV DLWFA flagship experiment only requires one measurement: the electron energy. Simulations predict 25 pC of electron charge with energies up to 100 GeV, so there should be adequate signal such that signal-to-noise ratio is not a concern. The ideal and most established electron energy measurement would employ a magnetic electron spectrometer. Electron spectrometers out to 80 GeV have already been demonstrated at SLAC. For this flagship, that capability would ideally be extended to 100 GeV, although meaningful measurements could be made even if the spectrometer only had 30-GeV capability. Alternative concepts are being developed to measure electron energies of this magnitude, such as methods based on stopping distances in materials, plasma wakefield acceleration, nuclear physics, Cherenkov emission, or the produced betatron spectrum.

**Experiment**

Campaigns will target the same goals as the 6-J and 30-J demonstrations: measure the extended focal volume, energy ramp up, acceleration optimization, and trapping control. The exact details of the plan will be informed based on lessons learned from the 6- and 30-J experiments and simulation work. If 100-GeV electron beams are successfully demonstrated, the flagship can be extended to investigate the role that radiation reaction plays at these electron energies [48]. This flagship can still be completed at reduced drive laser energies but at the cost of the maximum electron energy that can be produced.

## V: Conclusion

The ultimate goal of this flagship experiment on NSF OPAL using DLWFA is to demonstrate 100-GeV electron energies from a single plasma stage using a flying focus driven by the round NSF OPAL Alpha beam. An experimental path, and the computational and technical design work along that path, from the current status of the field to a single-stage, 100-GeV electron beam via DLWFA is outlined, and progress along that path is presented. There are three Grand Challenge technologies that need to be demonstrated: meter-scale flying-focus optics, a meter-scale plasma source, and a 100-GeV electron energy measurement.

LLE has developed the technologies required to fabricate the axiparabolas and echelons that produce an ultrafast flying focus. The all-optical demonstration of an ultrafast flying focus experimentally verifies that a controllable, ultrashort flying focus with a nearly transform- and diffraction-limited moving focus can be created over an extended focal region more than 50 Rayleigh ranges in length. Initial results from MTW-OPAL, the platform for the 6-J DLWFA demonstration experiment, show a tight, round focal spot over a 3.7 mm distance. New OSIRIS simulations of that platform indicate that using hydrogen for DLWFA provides a viable focus with limited refraction from ionization fronts.

We review the viable parameter space for DLWFA based on the scaling of its performance with laser and plasma parameters and compare it to traditional LWFA. These scalings indicate the need for ultrashort, high-energy laser technologies such as NSF OPAL to achieve groundbreaking electron energies using DLWFA. NSF OPAL could be the transformative technology that enables this next grand challenge in laser-plasma acceleration.

## Acknowledgements

This material is based upon work supported by the Department of Energy [National Nuclear Security Administration] University of Rochester "National Inertial Confinement Fusion Program" under Award Number(s) DE-NA0004144, by the U.S. National Science Foundation under Cooperative Agreement No. (PHY-2329970), and by Department of Energy Office of Fusion Energy under Award Number DE-SC0021057.

Disclaimer

This report was prepared as an account of work sponsored by an agency of the United States Government. Neither the United States Government nor any agency thereof, nor any of their employees, makes any warranty, express or implied, or assumes any legal liability or responsibility for the accuracy, completeness, or usefulness of any information, apparatus, product, or process disclosed, or represents that its use would not infringe privately owned rights. Reference herein to any specific commercial product, process, or service by trade name, trademark, manufacturer, or otherwise does not necessarily constitute or imply its endorsement, recommendation, or favoring by the United States Government or any agency thereof. The views and opinions of authors expressed herein do not necessarily state or reflect those of the United States Government or any agency thereof.

## Author Declarations

Conflict of Interest

The authors have no conflicts to disclose.

Author Contributions

J. L. Shaw: Conceptualization (equal); Data Curation (equal); Funding Acquisition (equal); Investigation (equal); Project Administration (equal); Resources (equal); Supervision (equal); Writing—Original Draft Preparation (lead); Writing—Review & Editing (lead). M. V. Ambat: Conceptualization (equal); Data Curation (equal); Formal Analysis (equal); Investigation (equal); Writing—Original Draft Preparation (supporting); Writing—Review & Editing (supporting). K. G. Miller: Conceptualization (equal); Data Curation (equal); Formal Analysis (equal); Investigation (equal); Methodology (equal); Resources (equal); Software (equal); Writing—Original Draft Preparation (supporting); Writing—Review & Editing (supporting). R. Boni: Conceptualization (equal); Investigation (equal); Methodology (equal). I. LaBelle: Investigation (supporting). W. B. Mori: Methodology (equal); Resources (equal); Software (equal); Supervision (equal). J. J. Pigeon: Investigation (equal); Supervision (equal). A. Rigatti: Investigation (equal). I. Settle: Investigation (supporting). L. Mack: Investigation (supporting). J. P. Palastro: Conceptualization (equal); Investigation (equal); Methodology (equal); Software (equal); Supervision (equal); Writing—Review & Editing (supporting). D. H. Froula: Conceptualization (equal); Funding Acquisition (equal); Investigation (equal); Methodology (equal); Project Administration (equal); Resources (equal); Supervision (equal); Writing—Original Draft Preparation (supporting); Writing—Review & Editing (supporting).

## Data Availability

The data that support the findings of this study are available from the corresponding author upon reasonable request.

# References

[1]  Executive Summary of 2023 Particle Physics Project Prioritization Panel (P5). https://www.usparticlephysics.org/2023-p5-report/executive-summary.
[2]  T. Tajima and J.M. Dawson. "Laser Electron Accelerator," Phys. Rev. Lett. **43, 267-270** (1979).
[3]  C. E. Clayton, K. A. Marsh, A. Dyson, M. Everett, A. Lal, W. P. Leemans, R. Williams, and C. Joshi. "Ultrahigh-Gradient Acceleration of Injected Electrons by Laser-Excited Relativistic Electron-Plasma Waves," Phys. Rev. Lett. **70**, 37-40 (1993).
[4]  M. Everett, A. Lal, D. Gordon, C.E. Clayton, K.A. Marsh, and C. Joshi. "Trapped Electron Acceleration by a Laser-Driven Relativistic Plasma-Wave," Nature **368,** 527-529 (1994).
[5]  C. G. R. Geddes, Cs. Toth, J. van Tilbog, E. Esarey, C. B. Schroeder, D. Bruhwiler, C. Nieter, J. Cary, and W. P. Leemans. "High-Quality Electron Beams from a Laser Wakefield Accelerator Using Plasma-Channel Guiding," Nature **431**, 538-541 (2004).
[6]  S. P. D. Mangles, C. D. Murphy, Z. Najmudin, A. G. R. Thomas, J. L. Collier, A. E. Dangor, E. J. Divall, P. S. Foster, J. G. Gallacher, C. J. Hooker *et al.* "Monoenergetic Beams of Relativistic Electrons from Intense Laser–Plasma Interactions," Nature **431**, 535-538 (2004).
[7]  J. Faure, Y. Glinec, A. Pukhov, S. Kiselev, S. Gordienko, E. Lefebvre, J.-P. Rousseau, F. Burgy, and V. Malka. *"*A Laser-Plasma Accelerator Producing Monoenergetic Electron Beams," Nature **431**, 541-544 (2004).
[8]  B. Miao, J. E. Shrock, L. Feder, R. C. Hollinger, J. Morrison, R. Nedbailo, A. Picksley, H. Song, S. Wang, J. J. Rocca *et al*. "Multi-GeV Electron Bunches from an All-Optical Laser Wakefield Accelerator" Phys Rev X, **12**(3), 031038 (2022).
[9]  A. J. Gonsalves, K. Nakamura, J. Daniels, C. Benedetti, C. Pieronek, T. C. H. de Raadt, S. Steinke, J. H. Bin, S. S. Bulanova, J. van Tilborg *et al.* "Petawatt Laser Guiding and Electron Beam Acceleration to 8 GeV in a Laser-Heated Capillary Discharge Waveguide," Phys. Rev. Lett. **122**, 8 (2019).
[10] C. Aniculaesei, T. Ha, S. Yoffe, L. Labun, S. Milton, E. McCary, M. M. Spinks, H. J. Quevedo, O. Z. Labun, R. Sain *et al*. "The acceleration of a high-charge electron bunch to 10 GeV in a 10-cm nanoparticle-assisted wakefield accelerator," Matter Radiat. Extremes **9**, 014001 (2024).
[11] A. Picksley, J. Stackhouse, C. Benedetti, K. Nakamura, H. E. Tsai, R. Li, B. Miao, J. E. Shrock, E. Rockafello, H. M. Milchberg *et al*., "Matched Guiding and Controlled Injection in Dark-Current-Free, 10-GeV-Class, Channel-Guided Laser-Plasma Accelerators," Phys. Rev. Lett. **133**, 255001 (2024).
[12] W. Leemans, and E. Esarey. "Laser-driven plasma-wave electron accelerators," Physics Today, **62**, 44-49 (2009).
[13] D. H. Froula, D. Turnbull, A. S. Davies, T. J. Kessler, D. Haberberger, J. P. Palastro, S.-W. Bahk, I. A. Begishev, R. Boni, S. Bucht *et al.* "Spatiotemporal control of laser intensity," Nature Photonics **12**, 5 (2018).
[14] J.P. Palastro, J.L. Shaw, P. Franke, D. Ramsey, T.T. Simpson, and D.H. Froula. "Dephasingless Laser Wakefield Acceleration," Phys. Rev. Lett. **124,** 13 (2020).
[15] J. Bromage, S.-W. Bahk, B. Barczys, A. Bolognesi, C. Dorrer, N. Ekanayake, C. Feng, E. Hill, C. Jeon, M. Krieger *et al.* "NSF OPAL: Laser System Design and Critical Technologies." *Bulletin of the American Physical Society* (2024).
[16] J. E. Ralph, K. A. Marsh, A. E. Pak, W. Lu, C. E. Clayton, F. Fang, W. B. Mori, and C. Joshi, "Self-Guiding of Ultrashort, Relativistically Intense Laser Pulses through Underdense Plasmas in the Blowout Regime," Phys. Rev. Lett. **102**, 175003 (2009).
[17] P. Sprangle and E. Esarey, "Interaction of ultrahigh laser fields with beams and plasmas," Phys. Fluids B **4**, 2241 (1992).
[18] C. G. Durfee and H. M. Milchberg, "Light pipe for high intensity laser pulses," Phys. Rev. Lett. **71**, 2409 (1993).

[19] W. Lu, M. Tzoufras, C. Joshi, F. S. Tsung, W. B. Mori, J. Vieira, R. A. Fonseca, and L. O. Silva. "Generating multi-GeV electron bunches using single stage laser wakefield acceleration in a 3D nonlinear regime," Phys. Rev. ST Accel. Beams **10**, 061301 (2007).

[20] K. G. Miller, J. R. Pierce, M. V. Ambat, J. L. Shaw, K. Weichman, W. B. More, D. H. Froula, and J. P. Palastro. "Dephasingless laser wakefield acceleration in the bubble regime," Sci. Rep. **13**, 21306 (2023). https://doi.org/10.1038/s41598-023-48249-4.

[21] *The International Linear Collider – Gateway to the Quantum Universe*. 2007; Available from: https://en.wikipedia.org/wiki/International_Linear_Collider#Cost.

[22] S. Smartsev, C. Caizergues, K. Oubrerie, J. Gautier, J.-P. Goddet, A. Tafzi, K. T. Phuoc, V. Malka, and C. Thaury. "Axiparabola: a long-focal-depth, high-resolution mirror for broadband high-intensity lasers," Opt. Lett. **44**, 3414-3417 (2019).

[23] P. Sprangle, B. Hafizi, J. R. Penano, R. F. Hubbard, A. Ting, A. Zigler, and T. M. Antonsen, Jr. "Stable Laser-Pulse Propagation in Plasma Chanels for GeV Electron Acceleration," Phys. Rev. Lett. **85**, 5110 (2000)

[24] P. Sprangle, J. R. Penano, B. Hafizi, R. F. Hubbard, A. Ting, D. F. Gordon, A. Zigler, and T. M. Antonsen, Jr. "GeV acceleration in tapered plasma channels," Phys. Plasmas **9**, 2364-2370 (2002).

[25] S. J. Yoon, J. P. Palastro, and H. M. Milchberg, "Quasi-Phase-Matched Laser Wakefield Acceleration," Phys. Rev. Lett. **112**, 134803 (2014).

[26] J. D. Sadler, C. Arran, H. Li, and K. A. Flippo. "Overcoming the dephasing limit in multiple-pulse laser wakefield acceleration," Phys. Rev. Accel. Beams **23**, 021303 (2020).

[27] A. Debus, R. Pausch, A. Huebl, K. Steiniger, R. Widera, T. E. Cowan, U. Schramm, and M. Bussmann. "Circumventing the Dephasing and Depletion Limits of Laser-Wakefield Acceleration," Phys. Rev. X **9**, 031044 (2019).

[28] C. Caizergues, S. Smartsev, V. Malka, and C. Thaury "Phase-locked laser-wakefield electron acceleration," Nat. Photonics **14**, 475 (2020).

[29] A. Liberman, R. Lahaye, S. Smartsev, S. Tata, S. Benracassa, A. Golovanov, E. Levine, C. Thaury, and V. Malka. "Use of spatiotemporal couplings and an axiparabola to control the velocity of peak intensity," Opt. Lett. **49**, 814-817 (2024).

[30] A. Liberman, A. Golovanov, S. Smartsev, S. Tata, I. A. Andriyash, S. Benracassa, E. Y. Levine, E. Kroupp, and V. Malka. "First Direct Observation of a Wakefield Generated with Structured Light," arXiv:2503.01516v1 [physics.acc-ph] 3 Mar 2025.

[31] C.B. Schroeder, F. Albert, C. Benedetti, J. Bromage, D. Bruhwiler, S. S. Bulanov, E. M. Campbell, N. M. Cook, B. Cros, M. C. Downer, *et al.*, "Linear colliders based on laser-plasma accelerators" JINST **18**, T06001(2023).

[32] https://www.clf.stfc.ac.uk/Pages/Vulcan-2020.aspx

[33] A. M. Maksimchuk, J. Nees, B. Hou, R. Anthony, F. Bayer, M. Burger, P. T. Campbell, G. Kalinchenko, S. R. Klein, Y. Ma *et al.* " Progress Report on the Commissioning of the ZEUS Facility." *Bulletin of the American Physical Society* (2024).

[34] J. H. Sung, H. W. Lee, J. Y. Yoo, J. W. Yoon, C. W. Lee, J. M. Yang, Y. J. Son, Y. H. Jang, S. K. Lee, and C. H. Nam, "4.2 PW, 20 fs Ti:sapphire laser at 0.1 Hz," Opt. Lett. **42**, 2058-2061 (2017).

[35] https://www.eli-beams.eu/facility/lasers/laser-3-hapls-1-pw-30-j-10-hz/

[36] J. J. Pigeon, P. Franke, M. Lim Pac Chong, J. Katz, R. Boni, C. Dorrer, J. P. Palastro, and D. H. Froula, "Ultrabroadband flying-focus using an axiparabola-echelon pair," Opt. Express **32**, 576-585 (2024).

[37] J. J. Pigeon, H. S. Markland, R. Boni, J. Kendrick, M. Lim Pac Chong, A. I. Elliott, K. G. Miller, J. P. Palastro, and D. H. Froula, "Measurements of THz produced by a two-color flying focus," *Submitted to 2025 Conference on Lasers and Electro-Optics, Optica Technical Digest.*

[38] M. V. Ambat, J. L. Shaw, J. J. Pigeon, K. G. Miller, T. T. Simpson, D. H. Froula, and J. P. Palastro. "Programmable-trajectory ultrafast flying focus pulses," Opt. Express **31**, 31354-31368 (2023).

[39] J.P. Palastro, K.G. Miller, M.R. Edwards, A.L. Elliott, L.S. Mack, D. Singh, and A.G.R. Thomas, "Arbitrary-velocity laser pulses in plasma waveguides," arXiv:2503.15690 [physics.plasm-ph] 19 Mar 2025.
[40] H. S. Markland, J. Rosenbluth, R. Boni, M.V. Ambat, C. Dorrer, J. P. Palastro, D. H. Froula, and J. J. Pigeon, "Measurements of a tunable-trajectory ultrabroadband flying-focus pulse using adaptive optics and an axiparabola," *Submitted to 2025 Conference on Lasers and Electro-Optics, Optica Technical Digest.*
[41] R. A. Fonseca, L. O. Silva, F. S. Tsung, V. K. Decyk, W. Lu , C. Ren, W. B. Mori, S. Deng, S. Lee, T. Katsouleas *et al.*, Lecture Notes in Computer Science, Vol. (Springer, Berlin, 2002), p. 342, Vol. 2331.
[42] J. R. Pierce, J. P. Palastro, F. Li, B. Malaca, D. Ramsey, J. Vieira, K. Weichman, and W. B. Mori, "Arbitrarily structured laser pulses," Phys. Rev. Research **5**, 013085 (2023).
[43] A. Pak, K. A. Marsh, S. F. Martins, W. Lu, W. B. Mori, and C. Joshi, Phys. Rev. Lett. **104**, 025003 (2010).
[44] C. McGuffey, A. G. R. Thomas, W. Schumaker, T. Matsuoka, V. Chvykov, F. J. Dollar, G. Kalintchenko, V. Yanovsky, A. Maksimchuk, V. Yu. Bychenkov *et al.*, Phys. Rev. Lett. **104**, 025004 (2010).
[45] J. Bromage, S.-W. Bahk, M. Bedzyk, I. A. Begishev, S. Bucht, C. Dorrer, C. Feng, C. Jeon, C. Mileham, R. G. Roides, K. Shaughnessy, M. J. Shoup III, M. Spilatro, B. Webb, D. Weiner, and J. D. Zuegel. "MTW-OPAL: a technology development platform for ultra-intense optical parametric chirped-pulse amplification systems," High Power Laser Science and Engineering **9**, 63 (2021) doi:10.1017/hpl.2021.45.
[46] M. V. Ambat, I. A. Settle, J. Shamlian, R. Boni, and J. L. Shaw, "Validation of a simple block model for permanent magnet electron spectrometers" *To be submitted to Review of Scientific Instruments*.
[47] F. Li, W. An, V. K. Decyk, X. Xu, M. J. Hogan, and W. B. Mori. "A quasi-static particle-in-cell algorithm based on an azimuthal Fourier decomposition for highly efficient simulations of plasma-based acceleration: QPAD," Computer Physics Communications **261**, 107784 (2021). https://doi.org/10.1016/j.cpc.2020.107784.
[48] A. A. Golovanov, E. N. Nerush, and I. Yu Kostyukov. "Radiation reaction-dominated regime of wakefield acceleration," New J. Physics **24**, 033011 (2022). https://doi.org/10.1088/1367-2630/ac53b9

# **Appendix: Derivation of Scalings for Figures 3 and 4**

To derive the scalings shown in Figure 3 and Figure 4, the following basic assumptions were made about the laser and plasma bubble:

$$\lambda_{laser} = 1.054 \text{ um}$$
$$k_p w_0 = k_p R = 2\sqrt{a_0}$$
$$\omega_p \tau_{laser} = \pi\sqrt{a_0}$$

Equation A1

The scaling laws from Ref. [19] for a traditional LWFA give:

$$L = \frac{2}{3}\left(\frac{\omega_0}{\omega_p}\right)^2 R \propto a_0^{1/2} n^{-3/2}$$
$$U_e = m_e c^2 \frac{k_p R}{2} \frac{k_p L}{2} \propto a_0 n^{-1}$$
$$U_{laser} = \frac{\pi}{2} I w_0^2 \tau_{laser} \propto a_0^{3/2} n^{-1/2}$$
$$N = \frac{(k_p R)^3}{30} \frac{4\pi\epsilon_0 m_e c^2}{k_p q_e^2} \propto a_0^{3/2} n^{-1/2}$$

Equation A2

$$\eta = \frac{NU_e}{U_{laser}} \propto a_0^{-1}$$

Here, L is the length of the accelerator (the dephasing length in this case), $U_e$ refers to the energy in one electron, $U_{laser}$ is the energy in the laser, N is the total number of electrons, and η is the efficiency.

In the case of a DLWFA, the length of the accelerator is not constrained by the dephasing length. It can be kept as a variable. The scaling lases for a DLWFA become

$$U_e = m_e c^2 \frac{k_p R}{2} \frac{k_p L}{2} \propto L a_0^{1/2} n^{1/2}$$
$$U_{laser} = \frac{\pi}{8} I w_0^2 \tau_{laser} \frac{\lambda_{laser} L}{\pi w_0^2} \propto L a_0^{5/2} n^{-1/2}$$
$$N = \frac{(k_p R)^3}{30} \frac{4\pi\epsilon_0 m_e c^2}{k_p q_e^2} \propto a_0^{3/2} n^{-1/2} \qquad \text{Equation A3}$$
$$\eta = \frac{NU_e}{U_{laser}} \propto n^{1/2} a_0^{-1/2} \propto \tau_{laser}^{-1}$$

These scalings make the same assumptions as the scalings for traditional LWFA, as defined in Eq. (A1).

All plots made using these scalings assume $a_0 = 2$. To create the scalings for an accelerator with *m* stages, the length of an individual stage is set to the total length divided by *m*. This means that the density increases by a factor of $m^{2/3}$, or approximately 22 for a 100-stage accelerator. The parameters $U_e$ and $U_{laser}$ then have to be multiplied by m to obtain their final value.